%% file: main.tex
\documentclass[a4paper,10pt]{IEEEtran}

\usepackage[T1]{fontenc}
\usepackage[latin9]{inputenc}
\usepackage{epsfig,psfrag}
\usepackage{amsmath,amsfonts,amssymb,amsxtra,bm}
\usepackage{amsthm}
\usepackage{graphicx}
\usepackage{color}
\usepackage{cite}
\usepackage{tabularx, enumerate}
\usepackage{multirow}
\usepackage{readarray}

\usepackage{flushend}

\usepackage[draft]{hyperref}
\hypersetup{bookmarks={false}}
\usepackage[table]{xcolor}

\newcolumntype{L}[1]{>{\raggedright\arraybackslash}p{#1}}
\newcolumntype{C}[1]{>{\centering\arraybackslash}p{#1}}
\newcolumntype{R}[1]{>{\raggedleft\arraybackslash}p{#1}}

\newcommand{\be}{\begin{equation}}
\newcommand{\ee}{\end{equation}}

\allowdisplaybreaks

\newcommand{\remove}[1]{}

\usepackage{tikz}

\usetikzlibrary{arrows,backgrounds,calc,positioning,shapes,shadows}
\usetikzlibrary{decorations.pathreplacing}
\usetikzlibrary{shapes.multipart, shapes.misc}
\usetikzlibrary{plotmarks}

\definecolor{txcircle}{RGB}{254, 233, 182} 
\definecolor{rxcircle}{RGB}{252, 76, 164} 

\definecolor{temporalgreenlight}{RGB}{194,224,194}
\definecolor{temporalgreen}{RGB}{0,128,0}
\definecolor{spatialred}{RGB}{255,0,0}
\definecolor{spatialredlight}{RGB}{255,194,194}
\definecolor{graylight}{RGB}{215,215,215}

\definecolor{txcircle}{RGB}{189, 193, 235} 
\definecolor{txtext}{RGB}{75, 88, 235} 

\definecolor{rxcircle}{RGB}{252, 167, 209} 
\definecolor{rxtext}{RGB}{252, 26, 114} 

\usepackage{pgfplotstable} 
\usepackage{pgfplots}
\usepgfplotslibrary{statistics}
\usetikzlibrary{pgfplots.statistics, math} 
  \pgfplotsset{compat=newest}
  \usetikzlibrary{plotmarks}
  \usepgfplotslibrary{patchplots}
  \usepackage{grffile}
  
  \makeatletter
  \pgfplotsset{
  	boxplot prepared from table/.code={
  		\def\tikz@plot@handler{\pgfplotsplothandlerboxplotprepared}%
  		\pgfplotsset{
  			/pgfplots/boxplot prepared from table/.cd,
  			#1,
  		}
  	},
  	/pgfplots/boxplot prepared from table/.cd,
  	table/.code={\pgfplotstablecopy{#1}\to\boxplot@datatable},
  	row/.initial=0,
  	make style readable from table/.style={
  		#1/.code={
  			\pgfplotstablegetelem{\pgfkeysvalueof{/pgfplots/boxplot prepared from table/row}}{##1}\of\boxplot@datatable
  			\pgfplotsset{boxplot/#1/.expand once={\pgfplotsretval}}
  		}
  	},
  	make style readable from table=lower whisker,
  	make style readable from table=upper whisker,
  	make style readable from table=lower quartile,
  	make style readable from table=upper quartile,
  	make style readable from table=median,
  	make style readable from table=lower notch,
  	make style readable from table=upper notch
  }
  \makeatother
  
  \newcommand{\pgfmathtruncategmacro}[2]{%
  	\pgfmathtruncatemacro\pgfmathresult{#2}%
  	\global\let#1\pgfmathresult
  }
\pgfplotsset{every axis/.append style={
                    label style={font=\footnotesize},
                    tick label style={font=\footnotesize},  
                    }}

\pgfplotsset{
	box plot width/.initial=1em,
	box plot/.style={
		/pgfplots/.cd,
		black,
		only marks,
		mark=-,
		mark size=\pgfkeysvalueof{/pgfplots/box plot width},
		/pgfplots/error bars/.cd,
		y dir=plus,
		y explicit,
	},
	box plot box/.style={
		/pgfplots/error bars/draw error bar/.code 2 args={densely dashed%
			\draw  ##1 -- ++(\pgfkeysvalueof{/pgfplots/box plot width},0pt) |- ##2 -- ++(-\pgfkeysvalueof{/pgfplots/box plot width},0pt) |- ##1 -- cycle;
		},
		/pgfplots/table/.cd,
		y index=2,
		y error expr={\thisrowno{3}-\thisrowno{2}},
		/pgfplots/box plot
	},
	box plot top whisker/.style={
		/pgfplots/error bars/draw error bar/.code 2 args={%
			\pgfkeysgetvalue{/pgfplots/error bars/error mark}%
			{\pgfplotserrorbarsmark}%
			\pgfkeysgetvalue{/pgfplots/error bars/error mark options}%
			{\pgfplotserrorbarsmarkopts}%
			\path ##1 -- ##2;
		},
		/pgfplots/table/.cd,
		y index=4,
		y error expr={\thisrowno{2}-\thisrowno{4}},
		/pgfplots/box plot
	},
	box plot bottom whisker/.style={
		/pgfplots/error bars/draw error bar/.code 2 args={%
			\pgfkeysgetvalue{/pgfplots/error bars/error mark}%
			{\pgfplotserrorbarsmark}%
			\pgfkeysgetvalue{/pgfplots/error bars/error mark options}%
			{\pgfplotserrorbarsmarkopts}%
			\path ##1 -- ##2;
		},
		/pgfplots/table/.cd,
		y index=5,
		y error expr={\thisrowno{3}-\thisrowno{5}},
		/pgfplots/box plot
	},
	box plot median/.style={densely dotted,
		/pgfplots/box plot
	}
}

\pgfplotsset{plot coordinates/math parser=false}
\newlength\figH
\newlength\figW
\usetikzlibrary{arrows.meta}

\usepackage{siunitx}

\usepackage[noend]{algpseudocode}
\usepackage[ruled, vlined, linesnumbered]{algorithm2e}
\algnewcommand\algorithmforeach{\textbf{for~each}}
\algdef{S}[FOR]{Foreach}[1]{\algorithmforeach\ #1\ \algorithmdo}
\def\centerarc[#1](#2)(#3:#4:#5)
    { \draw[#1] ($(#2)+({#5*cos(#3)},{#5*sin(#3)})$) arc (#3:#4:#5); }

\newcommand{\timeline}[7][] 
{
\draw[-, color=black!30] ($(#2) + (2,0)$) coordinate (A) -- ++(4.5,0);
	    \foreach \x in {0,1,...,#3}
	      {
	      \draw[->,#1] ($(A) + (#4+#5*\x,0)$) -- ++(0,0.3);
	      }
	    \coordinate (#7) at ($(A) + (#4+#5*#6,0)$);
	    \draw[#1] (#7) circle (.05cm);
}

\tikzset{cross/.style={cross out, draw=black, minimum size=2*(#1-\pgflinewidth), inner sep=0pt, outer sep=0pt},
cross/.default={0.5mm}}

\begin{document}

\title{Distance Estimation for BLE-based Contact Tracing -- A Measurement Study}

\author{\IEEEauthorblockN{ 
        Bernhard~Etzlinger, 
        Barbara~Nu\ss baumm\"uller,
        Philipp~Peterseil,
        and Karin~Anna~Hummel 
        \vspace*{3mm}}
        
	\IEEEauthorblockA {\small Johannes Kepler University, Linz, Austria, \{name.surname\}@jku.at\vspace*{.5mm}}
 \thanks{ 
 	This work has been supported by the LCM K2 Center within the framework of the Austrian COMET-K2 program.
 	}
	}
\vspace{-3mm}

\maketitle

\begin{abstract}   
 Mobile contact tracing apps are -- in principle -- a perfect aid to condemn the human-to-human spread of an infectious disease such as COVID-19 due to the wide use of smartphones worldwide. Yet, the unknown accuracy of contact estimation by wireless technologies hinders the broader use.
 We address this challenge by conducting a measurement study with a custom testbed to show the capabilities and limitations of Bluetooth Low Energy (BLE) in different scenarios. Distance estimation is based on interpreting the signal pathloss with a basic linear and a logarithmic model. Further, we compare our results with accurate ultra-wideband (UWB) distance measurements. 
 While the results 
indicate that 
distance estimation by BLE is not accurate enough, a contact detector can detect contacts below 2.5$\,$m with a true positive rate of 0.65 for the logarithmic and of 0.54 for the linear model. 
Further, the measurements reveal that multi-path signal propagation reduces the effect of body shielding and thus increases detection accuracy in indoor scenarios.
\end{abstract}


\maketitle

\input{introduction}

\input{relatedwork}

\input{implementation}

\input{experiments}

\input{conclusio}



\bibliographystyle{ieeetr_noParentheses}
\bibliography{main}

\end{document}

%% file: introduction.tex
\section{Introduction}
Contact tracing aims at fighting human-to-human infection spreading by identifying -- and isolating -- persons who were in close contact to an infected person.
Quarantine is one of the most effective measures to break infection chains~\cite{nature2020}. 
To support time-intense manual contact tracing, mobile contact tracing apps have been recently introduced that estimate and capture contacts as a measure against COVID-19.

Bluetooth Low Energy (BLE) is seen as the most promising wireless technology for contact tracing~\cite{martin2020demystifying}. 
The spatial distance between two smartphones is estimated based on
the received signal strength indicator (RSSI), which is known to be challenged by non line-of-sight (LOS) conditions such as shielding by human bodies and multi-path propagation in particular in indoor environments. It has been shown that BLE RSSI based distance estimation requires additional concepts to increase accuracy~\cite{leith2020bleRSSI}. 

Among alternative technologies, ultra-wideband (UWB) is a well-known technology often used for location estimation. Ultra-wideband (UWB)
based distance calculation uses time-of-flight (ToF) measures and may be 
leveraged for contact tracing.
Yet, UWB is currently only available in a few flagship smartphones such as Samsung Galaxy Note 20 or Apple iPhone 11. Further, UWB distance estimation relies on the cooperative exchange of time information~\cite{neirynck2016alternative}, which is a potential privacy threat and will consume additional energy. In our work, we will use UWB for comparison and consider it as ground-truth measurement system.

The focus of this paper rests on characterizing BLE RSSI based distance estimation for contact tracing.
We make the following contributions: 

\begin{itemize}
    \item We introduce our measurement testbed consisting of a mobile app that allows to capture BLE and UWB-based distance estimates, the latter for comparison. BLE-based distance estimation makes use of a linear and a logarithmic pathloss model that interpret BLE RSSI values measured onboard of the smartphone. UWB distance readings are retrieved from connected external UWB modules.
    
    \item We summarize the achieved accuracy of distance and exposure estimation in our measurement campaign in different scenarios consisting of 20'535 BLE logs. Our results confirm that using BLE RSSI for distance estimation is challenging 
    and that the linear distance estimation model provides better distance estimation accuracy than the logarithmic estimation model. We further show that in order to enhance exposure detection, awareness of the smartphones carrying position is more beneficial than knowledge about the environment.
    
    \item We study the effect of body shielding and multi-path propagation in an anechoic chamber (comparable to outdoor scenarios), LOS indoors, and in a corner scenario indoors with varying phone carrying positions. 
    We find that BLE multi-path propagation can 
    reduce body shielding effects. Finally, we compare BLE-based distance estimation with the more robust and accurate UWB ToF based estimation.
\end{itemize}


%% file: relatedwork.tex
\section{Related Work}
\label{sec:related}

BLE is widely used in mobile contact tracing apps currently promoted by health authorities worldwide. To unify and support contact tracing apps, the Google/Apple API for exposure notification based on BLE has been proposed for iOS\footnote{https://developer.apple.com/documentation/exposurenotification, accessed November 17th, 2020} and Android\footnote{https://developers.google.com/android/exposure-notifications/exposure-notifications-api, accessed November 17th, 2020} platforms. These APIs notify users about exposures to infected people
according to commonly agreed thresholds (too close, too long) while preserving privacy to a high degree. The Google/Apple notification API has been widely adopted in contact tracing apps~\cite{martin2020demystifying}. Yet, BLE-based distance estimation is error-prone. The study presented in~\cite{leith2020Plos} shows that staged contacts in a tram (less than 2\,m, longer than 15 minutes) do not lead to expected notifications of the API when using the contact tracing app of Switzerland or Germany, and only 50\% of exposures are detected by the Italian app. Thus, it is crucial to study and improve the effectiveness of BLE for contact tracing in real environments. 

The use of BLE for distance estimation and indoor localization has been thoroughly studied in the past, see, e.g.~\cite{benkic2008using,zhao2014does,touvat2014indoor,sadowski2018rssi}. These studies focus on communication of devices with no or little human interaction. As the contact tracing use case requires to capture realistic use scenarios of smartphones, those studies need to be extended.

Major contact tracing data-sets and data repositories are provided by the MIT PACT project~\cite{MITPACTdata2020} and the DP-3T initiative \cite{DP3Tdata2020}. Closest to the research presented in this paper is the study of BLE RSSI based distance estimation presented in~\cite{leith2020bleRSSI}, where several isolated effects such as body shielding or complex real-world situations are investigated.
In our study, we will add a new
dataset (joint effect of body shielding and multi-path), study model options to derive distance estimates based on BLE RSSI values, 
and extend the investigation by providing UWB-based measurements in addition to BLE RSSI logs.

%% file: implementation.tex

\section{Testbed Implementation}
\label{sec:implementation}

\begin{figure}
\centering
\begin{tabular}{cc}
\includegraphics[width=4cm]{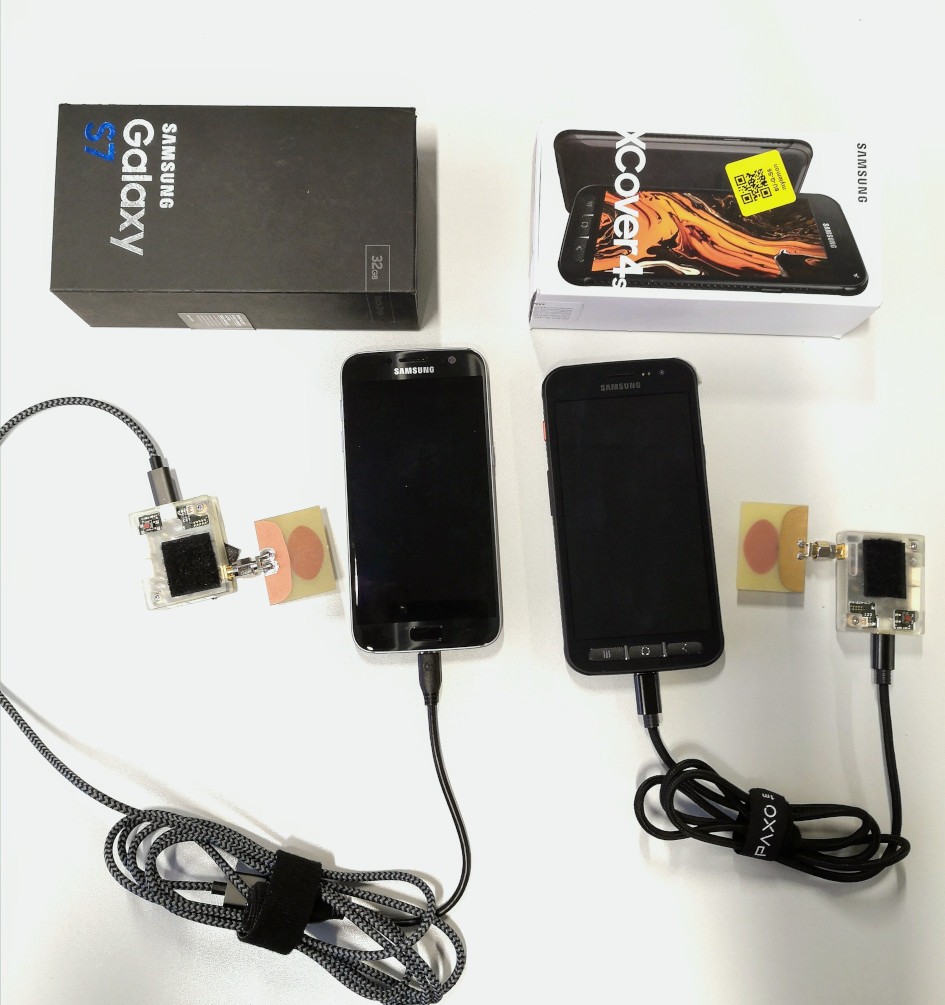} & 
\includegraphics[width=3.2cm]{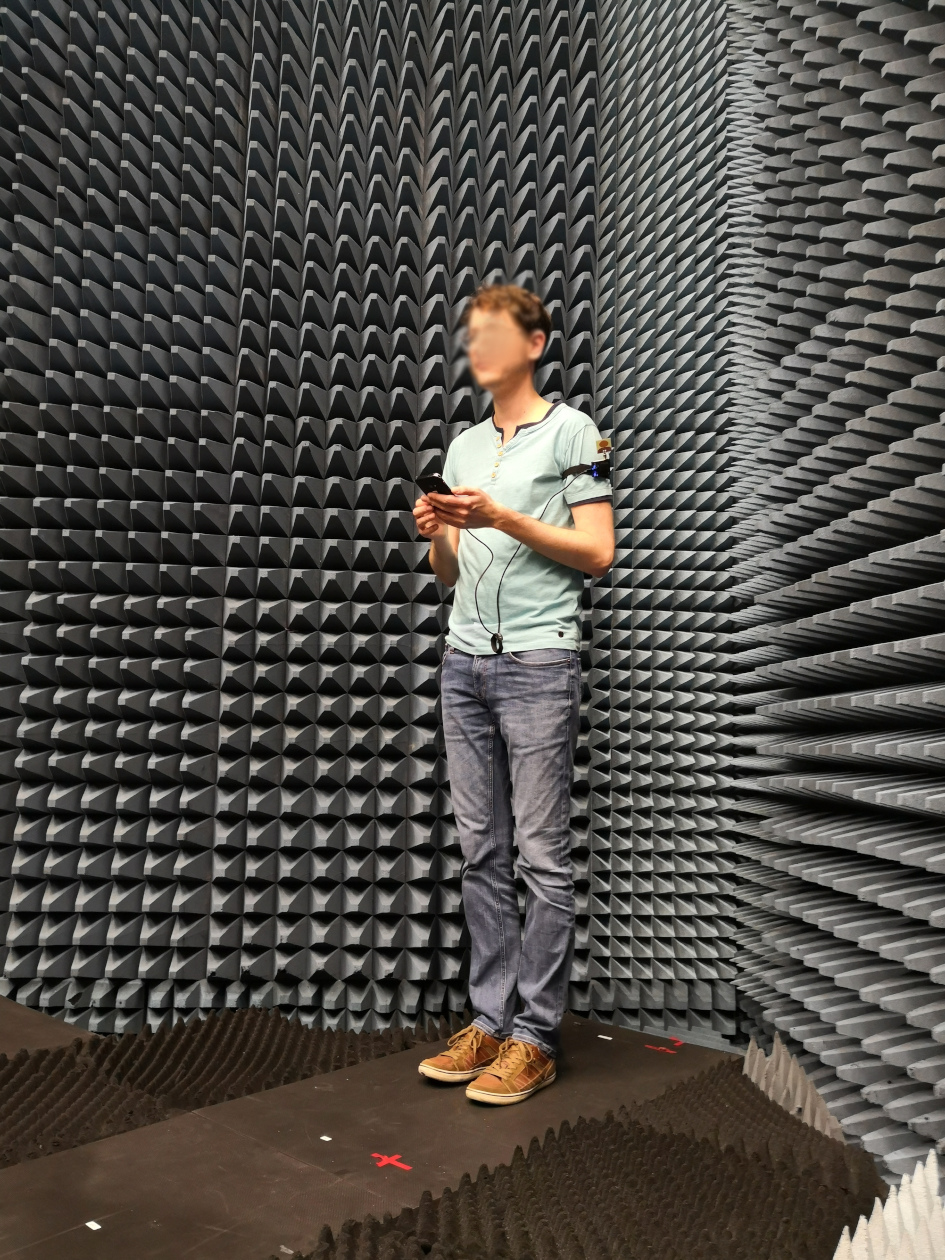} \\
\footnotesize{(a)} & \footnotesize{(b)}
\end{tabular}
\caption{Measurement equipment: (a) smartphones connected to UWB sensors, (b) UWB sensor mounted on the left arm of a person, smartphone held in front of trunk.}
\label{fig:testbed}
\end{figure}

The measurement testbed comprises off-the-shelf smartphones running a mobile measurement app that has been developed for the purpose of evaluating contact tracing technologies. The app captures BLE RSSI values provided by the onboard BLE module and UWB distance measurements retrieved from an external module. Fig.~\ref{fig:testbed} visualizes the testbed and its use.    

\subsection{Testbed Hardware and Software}

The testbed smartphones are of type Samsung Galaxy S7 
and Samsung XCover 4s 
(Android version 8 and 9, Android BLE API). We do not make use of the Google/Apple API as the aim of the study is to investigate lower-level information provided by the BLE module. 
Via USB port, the smartphones connect to the UWB sensors~\cite{etzlinger2020UWB}. The core component of the sensor is the Decawave DW1000 UWB transceiver chip, which is widely used for indoor localization.

The measurement app is a native Android app that provides a user interface to control an experiment and to log BLE RSSI and UWB distance values.
The app logs last recent scan values at configurable time intervals; 1\,s is the default setting.

Upon start, the app sends BLE advertisements based on the Generic Attribute Profile (GATT). When enabled by the user, a BLE scan is started resulting in an asynchronous return of newly scanned devices in BLE range. To restrict the devices reported in the scan, filtering by universally unique identifier (UUID) is employed.


\subsection{BLE RSSI Transformation}

BLE RSSI values are recorded on the smartphones as
obtained through the \textit{getRSSI()} Android function, which returns the received signal strength $P_{RX,i}$ in dBm, limited to the range of $[-127, 126]$\,dBm. Note that in the practical implementation, the lowest recorded RSSI value was -105$\,$dBm. From the RSSI values, the pathloss (PL) is calculated on the receiving device using calibration values from the GAEN database\footnote{https://developers.google.com/android/exposure-notifications/files/en-calibration-2020-08-12.csv} as
\begin{equation}
    \text{PL} = P_{\text{TX},j} - (P_{\text{RX},i} + \Delta_{\text{RX},i})\, , \label{eq:BLEcalibration}
\end{equation}
where $\Delta_{\text{RX},i}$ denotes the RX calibration value for scanning device $i$ and $P_{\text{TX},j}$ the calibrated TX power level of advertising device $j$. For the used Samsung XCover these are $P_{\text{TX}} = -24\,$dBm and $\Delta_{\text{RX}} = 5$\,dB, and for the Samsung S7 $P_{\text{TX}} = -33\,$dBm and $\Delta_{\text{RX}} = 10$\,dB.

\subsection{BLE Distance Estimation and Exposure Detection}
To detect exposures, we apply a distance based threshold detector. Thereby, we first obtain a model-based distance estimate $\hat{d}$ from a pathloss measurement. If the estimated distance is below the threshold distance, the data point is marked as positive, and otherwise negative. 

The model-based exposure estimation makes use of a linear pathloss model and a logarithmic pathloss model corresponding to known free space signal propagation properties.
The linear pathloss model is chosen to describe a basic correlation of RSSI values with ground truth distance and has no physical interpretation.
It is given by
\begin{equation}
	\text{PL}_\text{lin} = \text{PL}_\text{0,lin} + k\, d\, , \label{eq:model_lin}
\end{equation}
where $\text{PL}_\text{0,lin}$ in dB is the pathloss at distance $d=0$ and $k$ is the slope in dB/m.
The distance is then estimated by
\begin{equation}
  \hat{d}_\text{lin} = \text{max}\Big( \frac{\text{PL} - \text{PL}_\text{0,lin}}{k}, 0 \Big)\, ,\label{eq:estimation_lin}
\end{equation}
where the maximum function $\text{max}(\cdot,0)$ avoids the estimation of negative distances.
The log-distance pathloss model~\cite{benkic2008using} is given by
\begin{equation}
	\text{PL}_\text{ld} = \text{PL}_\text{0,ld} + 10\, \gamma \, \text{log}10\Big(\frac{d}{d_0}\Big) + X_g \, , \label{eq:model_ld}
\end{equation}
where  $\text{PL}_\text{0,ld}$ in dB is the pathloss at reference distance $d_0$, $\gamma$ is the pathloss exponent and $X_g$ is the zero-mean Gaussian noise in dB. In this work, we have chosen $d_0 = 2\,$m as it reflects the COVID-19 distance of interest. The estimated distance is then given by
\begin{equation}
 \hat{d}_\text{ld} = d_0 \, 10^{\frac{\text{PL} - \text{PL}_\text{0,ld}}{10\,\gamma}} \,. \label{eq:estimation_ld}
\end{equation}

\subsection{UWB Distance Estimation}

The UWB modules retrieve a distance estimate by time-of-flight (ToF) calculation, obtained from timestamps that are recorded upon packet transmission and reception. As the clocks of the receiver and the  transmitter are not synchronized, the ToF can only be estimated. 
The most common estimation approach
is
double-sided two-way ranging (DS-TWR)~\cite{neirynck2016alternative}, which requires TX and RX time stamps of three packet exchanges. 
We implement an extension to DS-TWR, known as cooperative synchronization and ranging~\cite{etzlinger2018synchronization}.
The distance estimates are updated every 250$\,$ms.

%% file: experiments.tex
\section{Measurement Study}
\label{sec:experiments}

In our measurement study, we aim at quantifying the accuracy of  our proposed distance estimators (linear model, logarithmic model) based on BLE RSSI values. As BLE signal propagation is known to be effected by body shielding and multi-path propagation~\cite{gezici2005localization}, we will in particular study these effects.
\subsection{Experiment Setup}

Two test persons are each equipped with a smartphone (Samsung xCover, Samsung Galaxy S7) and connected UWB sensor that is mounted on the left upper arm of the person. The experiments are recorded by the measurement app.
The following properties are varied in our experiments:

\noindent
\textbf{Carrying position:} The two test persons are always carrying the smartphone at the same position, which is either (i) head -- the smartphone is held at the left ear, 
(ii) trunk -- the smartphone is held in front of the trunk, 
or (iii) pelvis -- the smartphone is carried in the left front trouser's pocket.

\noindent
\textbf{Distance $d$:} The distance between the two persons is varied from 1$\,$m to 6$\,$m in steps of 
1$\,$m. 

\noindent
\textbf{Environment:} An anechoic chamber, a corridor, and a corner are selected, as depicted in Fig.~\ref{fig:scenarios} (a)-(c).

\noindent
\textbf{Orientation:} The relative orientation between the persons can be 
0$^{\circ}$, 90$^{\circ}$, 180$^{\circ}$, or 270$^{\circ}$ for head and pelvis carrying positions, and 0$^{\circ}$or 180$^{\circ}$ for trunk; see Fig.~\ref{fig:scenarios} (d).

\begin{figure*}
\centering
    \input{texfig/BLEalldata_fig}
    \vspace{-2mm}
    \readdef{texfig/DistErrorData/modelParamAll_logdist}{\mydatadef}
    \readArrayij{\mydatadef}{mydatald}{1}
    \readdef{texfig/DistErrorData/modelParamAll_lin}{\mydatadef}
    \readArrayij{\mydatadef}{mydatalin}{1}
    
    \caption{Heat map of BLE pathloss at six distances, grouped by three environments (corner, corridor, anechoic chamber). The model parameters of the linear model are: $\text{PL}_{0,\text{lin}}=\arrayij{mydatalin}{1}{1}\,$dB, $k=\arrayij{mydatalin}{2}{1}\,$dB/m with a correlation coefficient of $r = \arrayij{mydatalin}{3}{1}$. The parameters of the log-distance model are: $\text{PL}_{0,\text{ld}}=\arrayij{mydatald}{1}{1}\,$dB for reference distance $d_0=2.5\,$m, $\gamma=\arrayij{mydatald}{2}{1}$, $X_g$ noise standard deviation $\sigma = \arrayij{mydatald}{3}{1}\,$dB.}
    \vspace{-2mm}
    \label{fig:all_measurements}
\end{figure*}
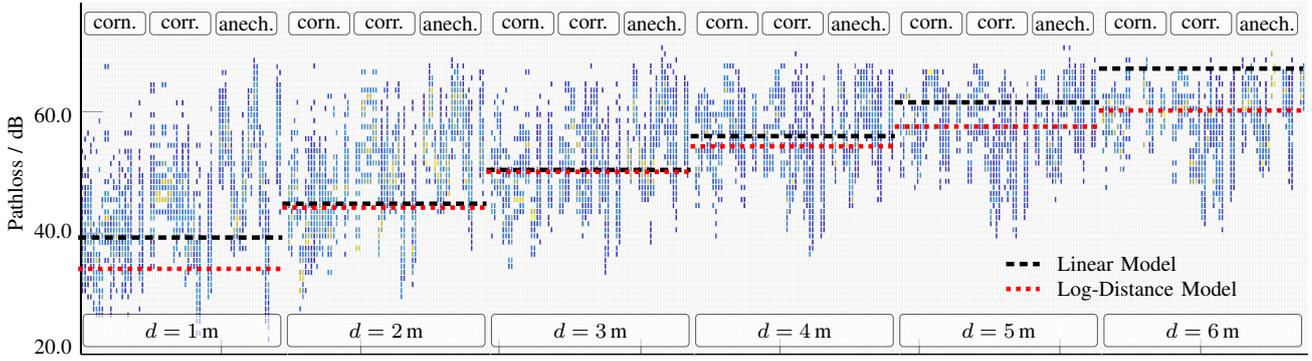

\vspace{0.1cm}
\noindent
 Overall, the experiment consists of 180 combinations, each setting is measured with a duration of 3\,min. 
 The BLE pathloss of the whole experiment is visualized in Fig.~\ref{fig:all_measurements} and ranges from 22\,dB to 70\,dB. In addition, the fitted linear model in \eqref{eq:model_lin}, here referred to as 'lin', and the fitted logarithmic model in \eqref{eq:model_ld}, referred to as 'l-d', are depicted. The correlation coefficient of the linear approximation is $r = 0.51$, which is a weak correlation between distance and pathloss caused by the high variability of pathloss at each distance. The noise $X_g$ standard deviation is high with $\sigma = 8.48$\,dB.

\begin{figure}
    \centering
    \input{texfig/scnarios}
    \vspace{-0.2cm}
    \caption{Schematics of environments: (a) anechoic chamber, (b) corridor, (c) corner. Locations of Person 1 and Person 2 are depicted by blue and red circles, respectively, at six timesteps (at each location, each person is $d/2$ away from 'x', the start location), and
    (d) relative orientations.}
    \label{fig:scenarios}
    \vspace{-2mm}
\end{figure}
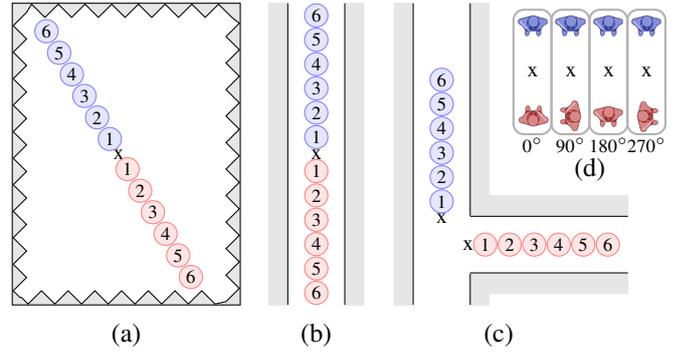

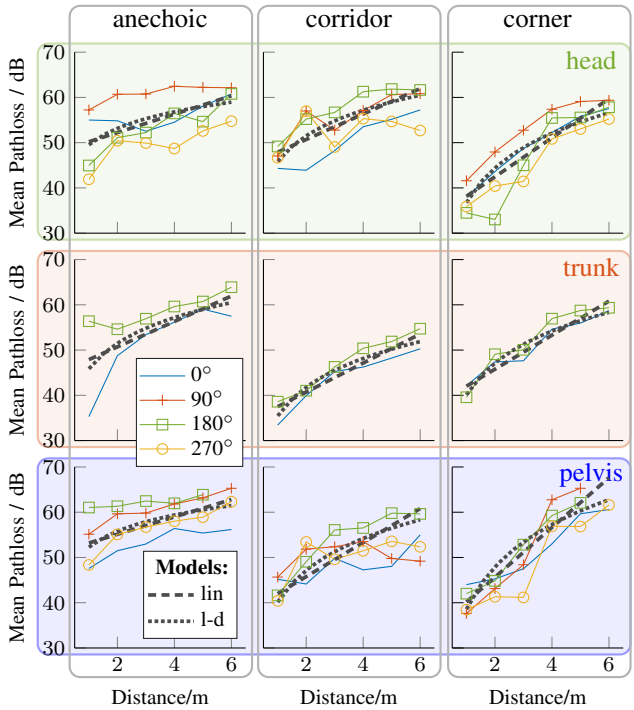
\begin{figure}[t]
	\centering
	\input{texfig/meanRSSI_distance_3x3.tex}
	\caption{BLE mean pathloss over actual distance 
	(mean std. deviation is 3.90$\,$dB). The black dashed and dotted lines depict the linear and, respectively, the log-distance fit, per environment and carrying position.}
	\label{fig:BLEresults}
	\vspace{-2mm}
\end{figure}

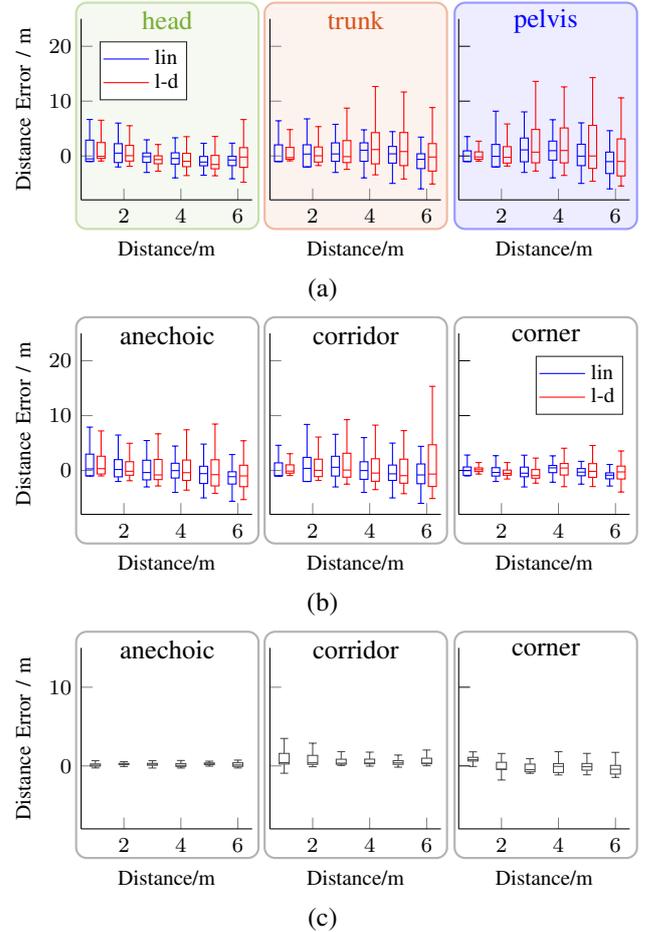
\begin{figure}[t]
	\centering
	\input{texfig/meanError_distancePos_1x3.tex}
	\\(a)\\[2mm]
	\input{texfig/meanError_distanceLoc_1x3.tex}
	\\(b)\\[2mm]
	\input{texfig/meanError_UWBdistanceLoc_1x3.tex}
	\\(c)
	\caption{Distance estimation error for (a) BLE with known carrying position and unknown environment, (b) BLE with unknown carrying position and known environment, and (c) UWB. The boxplots depict the median, 0.25 and 0.75 quantile and the corresponding whiskers.}
	\label{fig:boxplots}
	\vspace{-2mm}
\end{figure}

\subsection{Distance and Exposure Estimation Accuracy}

The distance estimation accuracy is described by the root-mean-square error (RMSE) of the distance estimate and the ground truth distance. The exposure detector is configured with a threshold of 2.5\,m, meaning that an exposure is detected when the estimated distance is below this threshold. The detector is evaluated by the true positive rate $r_p$, the true negative rate $r_n$, the $F_1$ score and the Matthews correlation coefficient (MCC).\footnote{The MCC is in the interval $[-1,1]$, where values >0 indicate a performance better than a random guess.}

Four scenarios are studied to assess how awareness of the environment and/or the carrying position influences estimation accuracy. For evaluation, the dataset is first divided along the known context settings (e.g., anechoic, corridor, corner), then each class is split equally into a training set and a test set. The training data is used to derive individual pathloss models for both model types (lin, l-d) considering censored pathloss measurements \cite{gustafson2015statistical}.


Comparing the case without context awareness (unknown/unknown) with the case of full awareness (known/known) in Tab.~\ref{tab:accuracy}, the detection rates $r_p$ and $r_n$ increase; hereby the impact of knowing the carrying position is larger than the impact of knowing the environment. This effect is also described by the MCC. Note that the MCC is always above zero, which indicates that the estimate is always better than a random guess. The highest observed accuracy is MCC$=0.47$ and $F_1=0.7$.

\begin{table}
	\caption{Accuracy of Exposure Estimation}
	\centering
	\setlength\tabcolsep{3pt} 
	\input{texfig/tab_summarizedvalues.tex}
	\label{tab:accuracy}
	\vspace{-2mm}
\end{table}

\subsection{Effect of Body Shielding}

To isolate the effect of body shielding, 
we now discuss the BLE attenuation in the anechoic chamber as visualized in Fig.~\ref{fig:BLEresults}, left column. 
The phone's carrying position and orientation have a major influence on the pathloss. In LOS scenarios (head, 180$^{\circ}$ and 270$^{\circ}$; trunk, 0$^{\circ}$; pelvis, 0$^{\circ}$) the mean pathloss is always lower than in the other non-LOS scenarios, at the respective same distance. The spread of pathloss is high in all carrying positions, i.e., up to 7.11\,dB (head, $d=1$\,m), 11.69\,dB (trunk, $d=1$\,m), and 6.53\,dB (pelvis, $d=1$\,m). The spread due to body shielding is often higher than the distance-dependent increase of attenuation. This is confirmed by the weak linear correlation coefficients between pathloss and distance (lin model), which are $r=0.4$ 
(head), $r=0.5$ (trunk), and $r=0.36$ 
(pelvis). (The l-d model shows a similar behavior.) 
These results show that in particular in short ranges important for contact tracing (up to 2m), accurate distance estimation cannot be expected.

The largest spread is found for head scenarios. For example, at a distance of 1\,m, the pathloss is 42$\,$dB and 44$\,$dB under LOS conditions (180$^{\circ}$ and 270$^{\circ}$) and when fully blocked by the head, it is 57$\,$dB (90$^{\circ}$). It is worth noting that with increasing distance this difference is decreasing due to a mean pathloss saturation at approximately 64$\,$dB (BLE packets with higher pathloss are lost).

\subsection{Effect of Multi-path Propagation}

Multi-path propagation is usually thought to hinder precise distance measurements. However, comparing the pathloss in a multi-path propagation environment (corridor, corner) to the pathloss in the anechoic chamber, the effect of body shielding is less severe due to reflections from the walls, as expressed by the smaller spread of pathloss at a given distance (see Fig.~\ref{fig:BLEresults}). The linear correlation coefficients between distance and pathloss (lin model) reflect this effect as well: $0.35 \leq r \leq 0.56$ (anechoic), $0.46 \leq r \leq 0.6$ (corridor), and $0.77 \leq r \leq 0.83$ (corner). These results indicate that in environments with multi-path signal propagation, distance estimation based on BLE pathloss may be possible without knowing the carrying position and orientation, which is  not the case for non multi-path environments such as outdoor environments. 

The highest correlation between distance and pathloss is observed in the corner scenario. At lower distances $d < 2$\,m, a low pathloss below 42\,dB is measured, which increases strongly with larger distances. The corner itself is a major cause of this behavior as at smaller distances LOS conditions are given. The corner obstructs the LOS path of the signal at larger distances.  

\subsection{Distance Estimation Error}

Fig.~\ref{fig:boxplots}(a) and (b) visualize the error statistics of BLE (lin and l-d models, individually fit to the scenario (head, trunk, pelvis and anechoic, corridor, corner)), and (c) the error statistics of UWB. For UWB, an overall distance RMSE of $0.9\,$m is observed, while BLE shows a RMSE of $3\,$m. This makes UWB a model technology w.r.t. its accuracy. Further, UWB may be used to collect ground truth measurements in realistic scenarios.

UWB measurements yield precise distance estimates for all experiment settings that allow LOS propagation. 
The signal of the LOS path is correctly recognized by the UWB transceiver amongst other reflected paths. Notably, in the anechoic chamber either  
the signal is strong enough to propagate through particular body parts or no signal is received at all. 

Wall penetration of the UWB signal may lead to small error-prone distance estimates (corridor, corner).
For example, see the negative distance errors in the corner environment. (Note that due to mounting the UWB sensors on the arm, position and orientation is not an issue.)  

Concerning BLE distance estimation errors, Fig.~\ref{fig:boxplots}(a) and (b) show that the spread of the distance errors increases with distance, irrespective of the context.  The l-d model also shows a larger spread of values than the lin model in most 
cases.

%% file: texfig/scnarios.tex
    \begin{tikzpicture}[]
    \begin{scope}
          \draw[fill=black!10] (0,0) rectangle (3,4);
          \draw[decorate, decoration=zigzag, fill=white] (0.1,0.1) rectangle (2.9,3.9);
          \coordinate (A) at (1.4,2) ;
          \node at (A) {\footnotesize x};

          \foreach \d in {1,...,6}
          {
          \node[draw=blue!50,circle, inner sep = 1pt, fill=blue!10] at  ($ (A) + (-\d/6,\d/3.5) - (-0.3/6,0.3/3.5) $) {\scriptsize \d};
          \node[draw=red!50,circle, inner sep = 1pt, fill=red!10] at  ($ (A) - (-\d/6,\d/3.5) + (-0.3/6,0.3/3.5) $) {\scriptsize \d};
          }
          \node at (1.5, -0.4) {(a)};
    \end{scope}
    \begin{scope}[xshift=3.25cm]
          \draw[draw=black!10, fill=black!10] (0.125,0) rectangle (1.375,4);
          \draw[draw=white, fill=white] (0.375,-0.1) rectangle (1.125,4.1);
          \draw (0.375,0) -- (0.375,4);
          \draw (1.125,0) -- (1.125,4);
          
          \coordinate (A) at (0.75,2) ;
          \node at (A) {\footnotesize x};

          \foreach \d in {1,...,6}
          {
          \node[draw=blue!50,circle, inner sep = 1pt, fill=blue!10] at  ($ (A) + (0,\d/3.1) - (0,0.3/3.1) $) {\scriptsize \d};
          \node[draw=red!50,circle, inner sep = 1pt, fill=red!10] at  ($ (A) - (0,\d/3.1) + (0,0.3/3.1) $) {\scriptsize \d};
          }
          \node at (0.75, -0.4) {(b)};
    \end{scope}
    \begin{scope}[xshift=4.9cm]
          \draw[draw=black!10, fill=black!10] (0.125,0) rectangle (1.375,4);
          \draw[draw=white, fill=white] (0.375,-0.1) rectangle (1.125,4.1);
          \draw[draw=black!10, fill=black!10] (1.375,0.15) rectangle (3.2,1.45);
          \draw (0.375,0) -- (0.375,4);
          \draw (1.125,0) -- (1.125,4);
          \draw[draw=white, fill=white] (1,0.425) rectangle (3.21,1.175);
          \draw (1.125,0.425) -- (3.2,0.425);
          \draw (1.125,1.175) -- (3.2,1.175);
          
          \coordinate (A) at (0.75,0.8) ;
          
          \node at ($(A) + (0.35,0)$) {\footnotesize x};
          \node at ($(A) + (0,0.35)$) {\footnotesize x};

          \foreach \d in {1,...,6}
          {
          \node[draw=blue!50,circle, inner sep = 1pt, fill=blue!10] at  ($ (A) + (0,\d/3.1) + (0,0.25)$) {\scriptsize \d};
          \node[draw=red!50,circle, inner sep = 1pt, fill=red!10] at  ($ (A) + (\d/3.1,0) + (0.25,0)$) {\scriptsize \d};
          }
          \node at (1.5, -0.4) {(c)};
    \end{scope}
    
    \begin{scope}[xshift=7.6cm, yshift=2.5cm]
    \node[draw=black!30, thick, rounded corners,
    minimum width=0.5cm,
    minimum height=1.65cm,anchor=south west] (person) at ($(-0.75,0) - (0.27,0.25)$) {};
    \node[inner sep=0pt, rotate=180] (personA) at (-0.75,0)
    {\includegraphics[width=0.4cm]{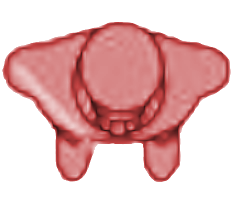}};
    \node [below of = personA, node distance=3.8mm] {\scriptsize 0$^{\circ}$};
    \node[above of = personA, node distance=12mm]
    {\includegraphics[width=0.4cm]{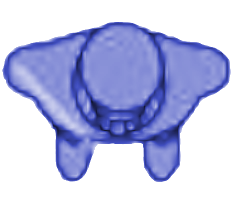}};
    \node[above of = personA, node distance=6mm]
    {\footnotesize x};
    
    \node[draw=black!30, thick, rounded corners,
    minimum width=0.5cm,
    minimum height=1.65cm,anchor=south west] (personA) at ($(-0.25,0) - (0.27,0.25)$) {};
    \node[inner sep=0pt, rotate=-90] (personA) at (-0.25,0)
    {\includegraphics[width=0.4cm]{fig/singlepersonTop_red.png}};
    \node [below of = personA, node distance=3.8mm] {\scriptsize 90$^{\circ}$};
    \node[above of = personA, node distance=12mm]
    {\includegraphics[width=0.4cm]{fig/singlepersonTop_blue.png}};
    \node[above of = personA, node distance=6mm]
    {\footnotesize x};
    
    \node[draw=black!30, thick, rounded corners,
    minimum width=0.5cm,
    minimum height=1.65cm,anchor=south west] (personA) at ($(0.25,0) - (0.27,0.25)$) {};
    \node[inner sep=0pt, rotate=0] (personA) at (0.25,0)
    {\includegraphics[width=0.4cm]{fig/singlepersonTop_red.png}};
    \node [below of = personA, node distance=3.8mm] {\scriptsize 180$^{\circ}$};
    \node[above of = personA, node distance=12mm]
    {\includegraphics[width=0.4cm]{fig/singlepersonTop_blue.png}};
    \node[above of = personA, node distance=6mm]
    {\footnotesize x};
    
    \node[draw=black!30, thick, rounded corners,
    minimum width=0.5cm,
    minimum height=1.65cm,anchor=south west] (personA) at ($(0.75,0) - (0.27,0.25)$) {};
    \node[inner sep=0pt, rotate=-270] (personA) at (0.75,0)
    {\includegraphics[width=0.4cm]{fig/singlepersonTop_red.png}};
    \node [below of = personA, node distance=3.8mm] {\scriptsize 270$^{\circ}$};
    \node[above of = personA, node distance=12mm]
    {\includegraphics[width=0.4cm]{fig/singlepersonTop_blue.png}};
    \node[above of = personA, node distance=6mm]
    {\footnotesize x};
    
    \node at (0,-0.7) {(d)};
    \end{scope}

    \end{tikzpicture}

%% file: texfig/BLEalldata_fig.tex
%
	
\begin{tikzpicture}

\begin{scope}[scale=1.1]
\node[draw=none,fill=none,anchor=south west] at (0,0){\includegraphics[height=5cm,width=3.19cm]{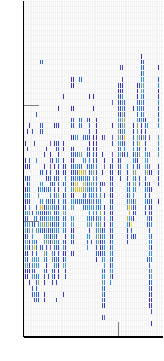}};
\node[draw=none,fill=none,anchor=south west] at (2.88,0){\includegraphics[height=5cm,width=2.69cm]{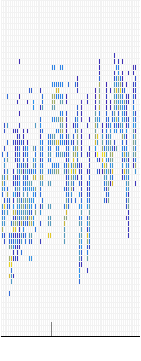}};
\node[draw=none,fill=none,anchor=south west] at (5.22,0){\includegraphics[height=5cm,width=2.81cm]{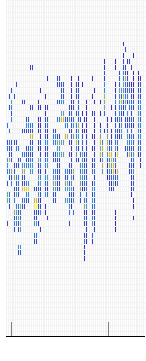}};
\node[draw=none,fill=none,anchor=south west] at (7.75,0){\includegraphics[height=5cm,width=2.72cm]{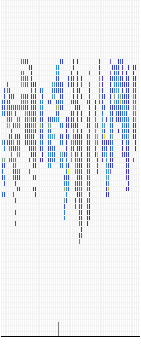}};
\node[draw=none,fill=none,anchor=south west] at (10.2,0){\includegraphics[height=5cm,width=2.69cm]{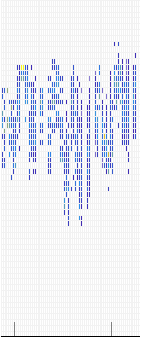}};
\node[draw=none,fill=none,anchor=south west] at (12.62,0){\includegraphics[height=5cm,width=2.74cm]{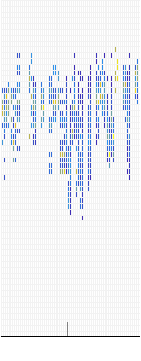}};

\foreach \y in {1,...,3} {
	\pgfmathsetmacro\result{\y * 20}
	\node at (0.2,1.4*\y-0.9) {\footnotesize \result};
}
\node [draw=none,fill=none, rotate=90] at (-0.25,2.6) {\footnotesize Pathloss / dB};


\readdef{texfig/DistErrorData/modelParamAll_lin}{\mydatadef}
\readArrayij{\mydatadef}{mydata1}{1}
\foreach \d in {1,...,6} {
	\pgfmathsetmacro\result{\d * \arrayij{mydata1}{1}{1} + \arrayij{mydata1}{2}{1}}
	\draw[densely dashed, line width = 1.7pt] ($(0.5,1.46*\result/20-1.05) + (\d*2.442-2.442,0)$) -- ($(0.5,1.46*\result/20-1.05) + (\d*2.442,0)$);
}


\readdef{texfig/DistErrorData/modelParamAll_logdist}{\mydatadef}
\readArrayij{\mydatadef}{mydata1}{1}
\foreach \d in {1,...,6} {
	\pgfmathsetmacro\result{log10( \d / 2.5 ) * \arrayij{mydata1}{2}{1} * 10 + \arrayij{mydata1}{1}{1}}
	\draw[dotted, line width = 1.7pt, draw=red] ($(0.5,1.46*\result/20-1.05) + (\d*2.442-2.442,0)$) -- ($(0.5,1.46*\result/20-1.05) + (\d*2.442,0)$);
}
\draw[densely dashed, line width = 1.7pt] (11.6,1.5) -- (12,1.5) coordinate (A);
\node[anchor=west] at ($(A) + (0.1,0)$) {\footnotesize Linear Model};
\draw[dotted, line width = 1.7pt, draw=red] (11.6,1.2) -- (12,1.2) coordinate (B);
\node[anchor=west] at ($(B) + (0.1,-0.02)$) {\footnotesize Log-Distance Model};

\foreach \y in {1,...,6} {
	\node [draw,fill=white,opacity=.6,text opacity=1,minimum width=2.6cm,rounded corners=.05cm] at (2.442*\y-0.7,0.7) {\footnotesize $d = \y\,$m};
	
		\node [draw,fill=white,opacity=.6,text opacity=1,minimum width=0.8cm,minimum height=0.3cm, inner sep = 0.03cm,rounded corners=.05cm] at (2.442*\y-1.5,4.4) {\footnotesize corn.};
		\node [draw,fill=white,opacity=.6,text opacity=1,minimum width=0.8cm,minimum height=0.3cm, inner sep = 0.03cm,rounded corners=.05cm] at (2.442*\y-0.72,4.4) {\footnotesize corr.};
		\node [draw,fill=white,opacity=.6,text opacity=1,minimum width=0.8cm,minimum height=0.3cm, inner sep = 0.03cm,rounded corners=.05cm] at (2.442*\y+0.07,4.4) {\footnotesize anech.};
}

\end{scope}
\end{tikzpicture}

%% file: texfig/meanRSSI_distance_3x3.tex
%
%
\definecolor{mycolor1}{rgb}{0.00000,0.44706,0.74118}%
\definecolor{mycolor2}{rgb}{0.85098,0.32549,0.09804}%
\definecolor{mycolor3}{rgb}{0.46667,0.67451,0.18824}%
\definecolor{mycolor4}{rgb}{0.92941,0.69412,0.12549}%
\begin{tikzpicture}

\node[draw=blue!40, thick, fill=blue!7, rounded corners,
    minimum width=7.8cm,
    minimum height=2.6cm,anchor=south west] (pelvis) at (-0.5,-0.1) {};
\node [draw=none] at (6.8cm,2.3cm) {\textcolor{blue}{pelvis}};

\node[draw=mycolor2!40, thick, fill=mycolor2!7, rounded corners,
    minimum width=7.8cm,
    minimum height=2.6cm,anchor=south west] (trunk) at (-0.5,2.65) {};
\node [draw=none] at (6.8cm,5.05cm) {\textcolor{mycolor2}{trunk}};

\node[draw=mycolor3!40, thick, fill=mycolor3!7, rounded corners,
    minimum width=7.8cm,
    minimum height=2.6cm,anchor=south west] (head) at (-0.5,5.4) {};
\node [draw=none] at (6.8cm,7.8cm) {\textcolor{mycolor3}{head}};

\node[draw=black!30, thick, rounded corners,
    minimum width=2.4cm,
    minimum height=8.9cm,anchor=south west] (pelvis) at (-0.08,-0.4) {};
\node[draw=black!30, thick, rounded corners,
    minimum width=2.4cm,
    minimum height=8.9cm,anchor=south west] (pelvis) at (2.4,-0.4) {};
\node[draw=black!30, thick, rounded corners,
    minimum width=2.4cm,
    minimum height=8.9cm,anchor=south west] (pelvis) at (4.9,-0.4) {};

\begin{axis}[%
width=2.25cm,
height=2.4cm,
at={(0cm,5.5cm)},
scale only axis,
xmin=0.5,
xmax=6.5,
ymin=30,
ymax=70,
ylabel style={font=\color{white!15!black}, font=\footnotesize},
ylabel={Mean Pathloss / dB},
title={anechoic},
axis x line*=bottom,
xticklabels=\empty,
axis y line*=left,
legend style={legend cell align=left, align=left, draw=white!15!black}
]
\addplot [color=mycolor1]
  table[row sep=crcr]{%
1	55\\
2	54.8315789473684\\
3	52.5193370165746\\
4	54.524861878453\\
5	58.0710382513661\\
6	60.7575757575758\\
};

\addplot [color=mycolor2, mark=+, mark options={solid, mycolor2}]
  table[row sep=crcr]{%
1	57.2266666666667\\
2	60.6890756302521\\
3	60.744\\
4	62.4528301886792\\
5	62.2142857142857\\
6	62.1304347826087\\
};

\addplot [color=mycolor3, mark=square, mark options={solid, mycolor3}]
  table[row sep=crcr]{%
1	44.9533678756477\\
2	51.103825136612\\
3	52.2265193370166\\
4	56.463687150838\\
5	54.6460674157303\\
6	60.7739726027397\\
};

\addplot [color=mycolor4, mark=o, mark options={solid, mycolor4}]
  table[row sep=crcr]{%
1	41.9044943820225\\
2	50.4277777777778\\
3	49.9175824175824\\
4	48.695652173913\\
5	52.5882352941176\\
6	54.7513812154696\\
};
\readdef{texfig/DistErrorData/modelParam11_lin}{\mydatadef}
\readArrayij{\mydatadef}{mydatalin}{1}

\addplot[domain=1:6, samples=2,color=black!70, densely dashed, line width =1.4pt]{\arrayij{mydatalin}{1}{1}*x+\arrayij{mydatalin}{2}{1}};

\readdef{texfig/DistErrorData/modelParam11_logdist}{\mydatadef}
\readArrayij{\mydatadef}{mydata}{1}

\addplot[domain=1:6, samples=10,color=black!70, densely dotted, line width =1.4pt]{(\arrayij{mydata}{1}{1} + 10*\arrayij{mydata}{2}{1}*log10(x/2)};

%

\end{axis}

\begin{axis}[%
width=2.25cm,
height=2.4cm,
at={(0cm,2.75cm)},
scale only axis,
xmin=0.5,
xmax=6.5,
ymin=30,
ymax=70,
ylabel style={font=\color{white!15!black}, font=\footnotesize},
ylabel={Mean Pathloss / dB},
axis x line*=bottom,
xticklabels=\empty,
axis y line*=left,
legend style={legend cell align=left, align=left, draw=white!15!black}
]
\addplot [color=mycolor1]
  table[row sep=crcr]{%
1	35.30078125\\
2	48.7578125\\
3	53.4083044982699\\
4	56.1666666666667\\
5	59.0495867768595\\
6	57.4715447154472\\
};

\addplot [color=mycolor3, mark=square, mark options={solid, mycolor3}]
  table[row sep=crcr]{%
1	56.3790322580645\\
2	54.604743083004\\
3	56.924\\
4	59.6244725738397\\
5	60.7235294117647\\
6	63.8979591836735\\
};

\readdef{texfig/DistErrorData/modelParam21_lin}{\mydatadef}
\readArrayij{\mydatadef}{mydatalin}{1}

\addplot[domain=1:6, samples=2,color=black!70, densely dashed, line width =1.4pt]{\arrayij{mydatalin}{1}{1}*x+\arrayij{mydatalin}{2}{1}};

\readdef{texfig/DistErrorData/modelParam21_logdist}{\mydatadef}
\readArrayij{\mydatadef}{mydata}{1}

\addplot[domain=1:6, samples=10,color=black!70, densely dotted, line width =1.4pt]{(\arrayij{mydata}{1}{1} + 10*\arrayij{mydata}{2}{1}*log10(x/2)};

%


\end{axis}

\begin{axis}[%
width=2.25cm,
height=2.4cm,
at={(0cm,0cm)},
scale only axis,
xmin=0.5,
xmax=6.5,
xlabel style={font=\color{white!15!black}, font=\footnotesize},
xlabel={Distance/m},
ymin=30,
ymax=70,
ylabel style={font=\color{white!15!black}, font=\footnotesize},
ylabel={Mean Pathloss / dB},
axis x line*=bottom,
axis y line*=left,
legend style={at={(0.4,0.05)}, anchor=south west, legend cell align=left, align=left, draw=white!15!black, font=\footnotesize, row sep=-1pt,inner sep=1pt}
]
\addlegendimage{empty legend}
\addlegendentry{\hspace{-.6cm}\textbf{Models:}}

\readdef{texfig/DistErrorData/modelParam31_lin}{\mydatadef}
\readArrayij{\mydatadef}{mydatalin}{1}

\addplot[domain=1:6, samples=2,color=black!70, densely dashed, line width =1.4pt]{\arrayij{mydatalin}{1}{1}*x+\arrayij{mydatalin}{2}{1}};
\addlegendentry{lin}

\readdef{texfig/DistErrorData/modelParam31_logdist}{\mydatadef}
\readArrayij{\mydatadef}{mydata}{1}

\addplot[domain=1:6, samples=10,color=black!70, densely dotted, line width =1.4pt]{(\arrayij{mydata}{1}{1} + 10*\arrayij{mydata}{2}{1}*log10(x/2)};
\addlegendentry{l-d}

\addplot [color=mycolor1]
  table[row sep=crcr]{%
1	47.5634920634921\\
2	51.4930555555556\\
3	52.9763779527559\\
4	56.403162055336\\
5	55.4016393442623\\
6	56.1793103448276\\
};

\addplot [color=mycolor2, mark=+, mark options={solid, mycolor2}]
  table[row sep=crcr]{%
1	55.1344537815126\\
2	59.6483516483516\\
3	59.812030075188\\
4	61.8181818181818\\
5	63.1774193548387\\
6	65.2857142857143\\
};

\addplot [color=mycolor3, mark=square, mark options={solid, mycolor3}]
  table[row sep=crcr]{%
1	61.0373831775701\\
2	61.3157894736842\\
3	62.4691358024691\\
4	61.962962962963\\
5	63.8684210526316\\
};

\addplot [color=mycolor4, mark=o, mark options={solid, mycolor4}]
  table[row sep=crcr]{%
1	48.3488372093023\\
2	55.1\\
3	56.7179487179487\\
4	58.021164021164\\
5	58.9075630252101\\
6	62.328125\\
};

%


\end{axis}

\begin{axis}[%
width=2.25cm,
height=2.4cm,
at={(2.484cm,5.5cm)},
scale only axis,
xmin=0.5,
xmax=6.5,
ymin=30,
ymax=70,
title={corridor},
axis x line*=bottom,
xticklabels=\empty,
axis y line*=left,
yticklabels=\empty,
legend style={legend cell align=left, align=left, draw=white!15!black}
]
\addplot [color=mycolor1]
  table[row sep=crcr]{%
1	44.3218390804598\\
2	43.9\\
3	48.2380952380952\\
4	53.5\\
5	55.1578947368421\\
6	57.2307692307692\\
};

\addplot [color=mycolor2, mark=+, mark options={solid, mycolor2}]
  table[row sep=crcr]{%
1	47.0476190476191\\
2	57\\
3	52.7727272727273\\
4	57.0909090909091\\
5	60.6111111111111\\
6	60.8235294117647\\
};

\addplot [color=mycolor3, mark=square, mark options={solid, mycolor3}]
  table[row sep=crcr]{%
1	49.1578947368421\\
2	55.2105263157895\\
3	56.6842105263158\\
4	61.2666666666667\\
5	61.8421052631579\\
6	61.6428571428571\\
};

\addplot [color=mycolor4, mark=o, mark options={solid, mycolor4}]
  table[row sep=crcr]{%
1	46.6818181818182\\
2	56.9444444444444\\
3	49.0476190476191\\
4	55.3\\
5	54.6875\\
6	52.6842105263158\\
};

\readdef{texfig/DistErrorData/modelParam21_lin}{\mydatadef}
\readArrayij{\mydatadef}{mydatalin}{1}

\addplot[domain=1:6, samples=2,color=black!70, densely dashed, line width =1.4pt]{\arrayij{mydatalin}{1}{1}*x+\arrayij{mydatalin}{2}{1}};

\readdef{texfig/DistErrorData/modelParam21_logdist}{\mydatadef}
\readArrayij{\mydatadef}{mydata}{1}

\addplot[domain=1:6, samples=10,color=black!70, densely dotted, line width =1.4pt]{(\arrayij{mydata}{1}{1} + 10*\arrayij{mydata}{2}{1}*log10(x/2)};

%

\end{axis}

\begin{axis}[%
width=2.25cm,
height=2.4cm,
at={(2.484cm,2.75cm)},
scale only axis,
xmin=0.5,
xmax=6.5,
ymin=30,
ymax=70,
axis x line*=bottom,
xticklabels=\empty,
axis y line*=left,
yticklabels=\empty,
legend style={legend cell align=left, align=left, draw=white!15!black}
]
\addplot [color=mycolor1]
  table[row sep=crcr]{%
1	33.4230769230769\\
2	39.9210526315789\\
3	45.2649572649573\\
4	46.2424242424242\\
5	48.2342342342342\\
6	50.2773109243698\\
};

\addplot [color=mycolor3, mark=square, mark options={solid, mycolor3}]
  table[row sep=crcr]{%
1	38.5955555555556\\
2	41.0253164556962\\
3	46.2735042735043\\
4	50.390350877193\\
5	51.8859649122807\\
6	54.7224669603524\\
};

\readdef{texfig/DistErrorData/modelParam22_lin}{\mydatadef}
\readArrayij{\mydatadef}{mydatalin}{1}

\addplot[domain=1:6, samples=2,color=black!70, densely dashed, line width =1.4pt]{\arrayij{mydatalin}{1}{1}*x+\arrayij{mydatalin}{2}{1}};

\readdef{texfig/DistErrorData/modelParam22_logdist}{\mydatadef}
\readArrayij{\mydatadef}{mydata}{1}

\addplot[domain=1:6, samples=10,color=black!70, densely dotted, line width =1.4pt]{(\arrayij{mydata}{1}{1} + 10*\arrayij{mydata}{2}{1}*log10(x/2)};

%

\end{axis}

\begin{axis}[%
width=2.25cm,
height=2.4cm,
at={(2.484cm,0cm)},
scale only axis,
xmin=0.5,
xmax=6.5,
xlabel style={font=\color{white!15!black}, font=\footnotesize},
xlabel={Distance/m},
ymin=30,
ymax=70,
axis x line*=bottom,
axis y line*=left,
yticklabels=\empty,
legend style={legend cell align=left, align=left, draw=white!15!black}
]
\addplot [color=mycolor1]
  table[row sep=crcr]{%
1	45.1774193548387\\
2	44.1376811594203\\
3	49.8167938931298\\
4	47.2619047619048\\
5	48.0393700787402\\
6	54.9915254237288\\
};

\addplot [color=mycolor2, mark=+, mark options={solid, mycolor2}]
  table[row sep=crcr]{%
1	45.669014084507\\
2	51.8387096774194\\
3	52.3740458015267\\
4	53.4957264957265\\
5	49.780303030303\\
6	49.1854304635762\\
};

\addplot [color=mycolor3, mark=square, mark options={solid, mycolor3}]
  table[row sep=crcr]{%
1	41.6172839506173\\
2	49.0538461538462\\
3	56.0793650793651\\
4	56.5254237288136\\
5	59.7934782608696\\
6	59.6283185840708\\
};

\addplot [color=mycolor4, mark=o, mark options={solid, mycolor4}]
  table[row sep=crcr]{%
1	40.4074074074074\\
2	53.4274193548387\\
3	49.5971223021583\\
4	51.4732824427481\\
5	53.5634920634921\\
6	52.3553719008264\\
};

\readdef{texfig/DistErrorData/modelParam23_lin}{\mydatadef}
\readArrayij{\mydatadef}{mydatalin}{1}

\addplot[domain=1:6, samples=2,color=black!70, densely dashed, line width =1.4pt]{\arrayij{mydatalin}{1}{1}*x+\arrayij{mydatalin}{2}{1}};

\readdef{texfig/DistErrorData/modelParam23_logdist}{\mydatadef}
\readArrayij{\mydatadef}{mydata}{1}

\addplot[domain=1:6, samples=10,color=black!70, densely dotted, line width =1.4pt]{(\arrayij{mydata}{1}{1} + 10*\arrayij{mydata}{2}{1}*log10(x/2)};

%

\end{axis}

\begin{axis}[%
width=2.25cm,
height=2.4cm,
at={(4.968cm,5.5cm)},
scale only axis,
xmin=0.5,
xmax=6.5,
every outer y axis line/.append style={black},
every y tick label/.append style={font=\color{black}},
every y tick/.append style={black},
ymin=30,
ymax=70,
title={corner},
axis x line*=bottom,
xticklabels=\empty,
axis y line*=left,
yticklabels=\empty,
legend style={legend cell align=left, align=left, draw=white!15!black}
]
\addplot [color=mycolor1]
  table[row sep=crcr]{%
1	37.9930555555556\\
2	43.6206896551724\\
3	48.7045454545455\\
4	52.2368421052632\\
5	55.8157894736842\\
6	57.5277777777778\\
};

\addplot [color=mycolor2, mark=+, mark options={solid, mycolor2}]
  table[row sep=crcr]{%
1	41.5869565217391\\
2	47.9318181818182\\
3	52.75\\
4	57.3823529411765\\
5	59.0909090909091\\
6	59.3703703703704\\
};

\addplot [color=mycolor3, mark=square, mark options={solid, mycolor3}]
  table[row sep=crcr]{%
1	34.4814814814815\\
2	33.0217391304348\\
3	45\\
4	55.4545454545455\\
5	55.5652173913044\\
6	57.8235294117647\\
};

\addplot [color=mycolor4, mark=o, mark options={solid, mycolor4}]
  table[row sep=crcr]{%
1	35.8723404255319\\
2	40.4142857142857\\
3	41.4324324324324\\
4	50.8333333333333\\
5	53.0294117647059\\
6	55.2222222222222\\
};

\readdef{texfig/DistErrorData/modelParam13_lin}{\mydatadef}
\readArrayij{\mydatadef}{mydatalin}{1}

\addplot[domain=1:6, samples=2,color=black!70, densely dashed, line width =1.4pt]{\arrayij{mydatalin}{1}{1}*x+\arrayij{mydatalin}{2}{1}};

\readdef{texfig/DistErrorData/modelParam13_logdist}{\mydatadef}
\readArrayij{\mydatadef}{mydata}{1}

\addplot[domain=1:6, samples=10,color=black!70, densely dotted, line width =1.4pt]{(\arrayij{mydata}{1}{1} + 10*\arrayij{mydata}{2}{1}*log10(x/2)};

%

\end{axis}

\begin{axis}[%
width=2.25cm,
height=2.4cm,
at={(4.968cm,2.75cm)},
scale only axis,
xmin=0.5,
xmax=6.5,
every outer y axis line/.append style={black},
every y tick label/.append style={font=\color{black}},
every y tick/.append style={black},
ymin=30,
ymax=70,
axis x line*=bottom,
xticklabels=\empty,
axis y line*=left,
yticklabels=\empty,
legend style={legend cell align=left, align=left, draw=white!15!black}
]
\addplot [color=mycolor1]
  table[row sep=crcr]{%
1	42.1020408163265\\
2	47.3783783783784\\
3	47.6356589147287\\
4	54.6238532110092\\
5	55.9523809523809\\
6	58.7\\
};

\addplot [color=mycolor3, mark=square, mark options={solid, mycolor3}]
  table[row sep=crcr]{%
1	39.568\\
2	49.0689655172414\\
3	50.0687022900763\\
4	56.94\\
5	58.7361111111111\\
6	59.3461538461538\\
};

\readdef{texfig/DistErrorData/modelParam23_lin}{\mydatadef}
\readArrayij{\mydatadef}{mydatalin}{1}

\addplot[domain=1:6, samples=2,color=black!70, densely dashed, line width =1.4pt]{\arrayij{mydatalin}{1}{1}*x+\arrayij{mydatalin}{2}{1}};

\readdef{texfig/DistErrorData/modelParam23_logdist}{\mydatadef}
\readArrayij{\mydatadef}{mydata}{1}

\addplot[domain=1:6, samples=10,color=black!70, densely dotted, line width =1.4pt]{(\arrayij{mydata}{1}{1} + 10*\arrayij{mydata}{2}{1}*log10(x/2)};

%

\end{axis}

\begin{axis}[%
width=2.25cm,
height=2.4cm,
at={(4.968cm,0cm)},
scale only axis,
xmin=0.5,
xmax=6.5,
xlabel style={font=\color{white!15!black}, font=\footnotesize},
xlabel={Distance/m},
every outer y axis line/.append style={black},
every y tick label/.append style={font=\color{black}},
every y tick/.append style={black},
ymin=30,
ymax=70,
axis x line*=bottom,
axis y line*=left,
yticklabels=\empty,
legend style={at={(-1.85,1.0)}, anchor=south west, legend cell align=left, align=left, draw=white!15!black, font=\footnotesize, row sep=-2pt,inner sep=1pt}
]

\addplot [color=mycolor1]
  table[row sep=crcr]{%
1	44\\
2	45.2808510638298\\
3	47.5\\
4	53\\
5	59.6567164179104\\
6	60.6\\
};
\addlegendentry{0$^{\circ}$}

\addplot [color=mycolor2, mark=+, mark options={solid, mycolor2}]
  table[row sep=crcr]{%
1	37.6422764227642\\
2	43.1256544502618\\
3	48.4\\
4	62.75\\
5	65.2857142857143\\
};
\addlegendentry{90$^{\circ}$}

\addplot [color=mycolor3, mark=square, mark options={solid, mycolor3}]
  table[row sep=crcr]{%
1	42\\
2	44.8203125\\
3	52.8260869565217\\
4	59.25\\
5	62.0689655172414\\
};
\addlegendentry{180$^{\circ}$}

\addplot [color=mycolor4, mark=o, mark options={solid, mycolor4}]
  table[row sep=crcr]{%
1	38.469387755102\\
2	41.3214285714286\\
3	41.1739130434783\\
4	56.8936170212766\\
5	56.8347826086957\\
6	61.6\\
};
\addlegendentry{270$^{\circ}$}


\readdef{texfig/DistErrorData/modelParam33_lin}{\mydatadef}
\readArrayij{\mydatadef}{mydatalin}{1}

\addplot[domain=1:6, samples=2,color=black!70, densely dashed, line width =1.4pt]{\arrayij{mydatalin}{1}{1}*x+\arrayij{mydatalin}{2}{1}};

\readdef{texfig/DistErrorData/modelParam33_logdist}{\mydatadef}
\readArrayij{\mydatadef}{mydata}{1}

\addplot[domain=1:6, samples=10,color=black!70, densely dotted, line width =1.4pt]{(\arrayij{mydata}{1}{1} + 10*\arrayij{mydata}{2}{1}*log10(x/2)};



\end{axis}
\end{tikzpicture}%

%% file: texfig/meanError_distancePos_1x3.tex
%
%
\definecolor{mycolor1}{rgb}{0.00000,0.44706,0.74118}%
\definecolor{mycolor2}{rgb}{0.85098,0.32549,0.09804}%
\definecolor{mycolor3}{rgb}{0.46667,0.67451,0.18824}%
\definecolor{mycolor4}{rgb}{0.92941,0.69412,0.12549}%
\begin{tikzpicture}

%
%

\node[draw=mycolor3!40, fill=mycolor3!7, thick, rounded corners,
    minimum width=2.4cm,
    minimum height=3.0cm,anchor=south west] (pelvis) at (-0.08,-0.4) {};
\node at ($(pelvis.north) + (0,-0.25cm)$) {\textcolor{mycolor3}{head}};

\node[draw=mycolor2!40, thick, fill=mycolor2!7, rounded corners,
    minimum width=2.4cm,
    minimum height=3.0cm,anchor=south west] (pelvis) at (2.4,-0.4) {};
\node at ($(pelvis.north) + (0,-0.25cm)$) {\textcolor{mycolor2}{trunk}};

\node[draw=blue!40, thick, fill=blue!7, rounded corners,
    minimum width=2.4cm,
    minimum height=3.0cm,anchor=south west] (pelvis) at (4.9,-0.4) {};
\node at ($(pelvis.north) + (0,-0.25cm)$) {\textcolor{blue}{pelvis}};

\begin{axis}[%
boxplot/draw direction=y,
width=2.25cm,
height=2.4cm,
at={(0cm,0cm)},
scale only axis,
xmin=0.5,
xmax=6.5,
xlabel style={font=\color{white!15!black}, font=\footnotesize},
xlabel={Distance/m},
ymin=-8,
ymax=25,
ylabel style={font=\color{white!15!black}, font=\footnotesize},
ylabel={Distance Error / m},
axis x line*=bottom,
axis y line*=left,
legend style={legend cell align=left, align=left, draw=white!15!black}
]

%

\pgfplotstableread{texfig/DistErrorData/errorBoxPlotPos1_lin.tex}\boxdatalin
\pgfplotsinvokeforeach{0,...,5}{
	\addplot+[
	boxplot prepared from table={
		table=\boxdatalin,
		row=#1,
		lower whisker=lw,
		upper whisker=uw,
		lower quartile=lq,
		upper quartile=uq,
		median=med}, 
	boxplot prepared={
		draw position={0.8 + \plotnumofactualtype},
		box extend=0.28}, color=blue, solid,
	]
	coordinates {};
}

\pgfplotstableread{texfig/DistErrorData/errorBoxPlotPos1_logdist.tex}\boxdatalogdist
\pgfplotsinvokeforeach{0,...,5}{
	\addplot+[
	boxplot prepared from table={
		table=\boxdatalogdist,
		row=#1,
		lower whisker=lw,
		upper whisker=uw,
		lower quartile=lq,
		upper quartile=uq,
		median=med}, 
	boxplot prepared={
		draw position={1.2 + \plotnumofactualtype-6},
		box extend=0.28}, color=red, solid,
	]
	coordinates {};
}

%
\end{axis}

\begin{axis}[%
boxplot/draw direction=y,
width=2.25cm,
height=2.4cm,
at={(2.484cm,0cm)},
scale only axis,
xmin=0.5,
xmax=6.5,
xlabel style={font=\color{white!15!black}, font=\footnotesize},
xlabel={Distance/m},
ymin=-8,
ymax=25,
axis x line*=bottom,
axis y line*=left,
yticklabels=\empty,
legend style={legend cell align=left, align=left, draw=white!15!black}
]

\pgfplotstableread{texfig/DistErrorData/errorBoxPlotPos2_lin.tex}\boxdatalin
\pgfplotsinvokeforeach{0,...,5}{
	\addplot+[
	boxplot prepared from table={
		table=\boxdatalin,
		row=#1,
		lower whisker=lw,
		upper whisker=uw,
		lower quartile=lq,
		upper quartile=uq,
		median=med}, 
	boxplot prepared={
		draw position={0.8 + \plotnumofactualtype},
		box extend=0.28}, color=blue, solid,
	]
	coordinates {};
}

\pgfplotstableread{texfig/DistErrorData/errorBoxPlotPos2_logdist.tex}\boxdatalogdist
\pgfplotsinvokeforeach{0,...,5}{
	\addplot+[
	boxplot prepared from table={
		table=\boxdatalogdist,
		row=#1,
		lower whisker=lw,
		upper whisker=uw,
		lower quartile=lq,
		upper quartile=uq,
		median=med}, 
	boxplot prepared={
		draw position={1.2 + \plotnumofactualtype-6},
		box extend=0.28}, color=red, solid,
	]
	coordinates {};
}

%

%

\end{axis}

\begin{axis}[%
boxplot/draw direction=y,
width=2.25cm,
height=2.4cm,
at={(4.968cm,0cm)},
scale only axis,
xmin=0.5,
xmax=6.5,
xlabel style={font=\color{white!15!black}, font=\footnotesize},
xlabel={Distance/m},
every outer y axis line/.append style={black},
every y tick label/.append style={font=\color{black}},
every y tick/.append style={black},
ymin=-8,
ymax=25,
axis x line*=bottom,
axis y line*=left,
yticklabels=\empty,
legend style={at={(-2.1,0.55)}, anchor=south west, legend cell align=left, align=left, draw=white!15!black, font=\footnotesize, row sep=-2pt,inner sep=1pt}
]

\addlegendimage{no markers,blue}
\addlegendentry{lin}

\addlegendimage{no markers,red}
\addlegendentry{l-d}

\pgfplotstableread{texfig/DistErrorData/errorBoxPlotPos3_lin.tex}\boxdatalin
\pgfplotsinvokeforeach{0,...,5}{
	\addplot+[
	boxplot prepared from table={
		table=\boxdatalin,
		row=#1,
		lower whisker=lw,
		upper whisker=uw,
		lower quartile=lq,
		upper quartile=uq,
		median=med}, 
	boxplot prepared={
		draw position={0.8 + \plotnumofactualtype},
		box extend=0.28}, color=blue, solid,
	]
	coordinates {};
}

\pgfplotstableread{texfig/DistErrorData/errorBoxPlotPos3_logdist.tex}\boxdatalogdist
\pgfplotsinvokeforeach{0,...,5}{
	\addplot+[
	boxplot prepared from table={
		table=\boxdatalogdist,
		row=#1,
		lower whisker=lw,
		upper whisker=uw,
		lower quartile=lq,
		upper quartile=uq,
		median=med}, 
	boxplot prepared={
		draw position={1.2 + \plotnumofactualtype-6},
		box extend=0.28}, color=red, solid,
	]
	coordinates {};
}
%

%

\end{axis}
\end{tikzpicture}%

%% file: texfig/meanError_distanceLoc_1x3.tex
%
%
\definecolor{mycolor1}{rgb}{0.00000,0.44706,0.74118}%
\definecolor{mycolor2}{rgb}{0.85098,0.32549,0.09804}%
\definecolor{mycolor3}{rgb}{0.46667,0.67451,0.18824}%
\definecolor{mycolor4}{rgb}{0.92941,0.69412,0.12549}%
\begin{tikzpicture}

%
%

\node[draw=black!30, thick, rounded corners,
    minimum width=2.4cm,
    minimum height=3.0cm,anchor=south west] (pelvis) at (-0.08,-0.4) {};
\node at ($(pelvis.north) + (0,-0.25cm)$) {anechoic};

\node[draw=black!30, thick, rounded corners,
    minimum width=2.4cm,
    minimum height=3.0cm,anchor=south west] (pelvis) at (2.4,-0.4) {};
\node at ($(pelvis.north) + (0,-0.25cm)$) {corridor};

\node[draw=black!30, thick, rounded corners,
    minimum width=2.4cm,
    minimum height=3.0cm,anchor=south west] (pelvis) at (4.9,-0.4) {};
\node at ($(pelvis.north) + (0,-0.25cm)$) {corner};

\begin{axis}[%
boxplot/draw direction=y,
width=2.25cm,
height=2.4cm,
at={(0cm,0cm)},
scale only axis,
xmin=0.5,
xmax=6.5,
xlabel style={font=\color{white!15!black}, font=\footnotesize},
xlabel={Distance/m},
ymin=-8,
ymax=25,
ylabel style={font=\color{white!15!black}, font=\footnotesize},
ylabel={Distance Error / m},
axis x line*=bottom,
axis y line*=left,
legend style={legend cell align=left, align=left, draw=white!15!black}
]

\pgfplotstableread{texfig/DistErrorData/errorBoxPlotLoc1_lin.tex}\boxdatalin
\pgfplotsinvokeforeach{0,...,5}{
	\addplot+[
	boxplot prepared from table={
		table=\boxdatalin,
		row=#1,
		lower whisker=lw,
		upper whisker=uw,
		lower quartile=lq,
		upper quartile=uq,
		median=med}, 
	boxplot prepared={
		draw position={0.8 + \plotnumofactualtype},
		box extend=0.28}, color=blue, solid,
	]
	coordinates {};
}

\pgfplotstableread{texfig/DistErrorData/errorBoxPlotLoc1_logdist.tex}\boxdatalogdist
\pgfplotsinvokeforeach{0,...,5}{
	\addplot+[
	boxplot prepared from table={
		table=\boxdatalogdist,
		row=#1,
		lower whisker=lw,
		upper whisker=uw,
		lower quartile=lq,
		upper quartile=uq,
		median=med}, 
	boxplot prepared={
		draw position={1.2 + \plotnumofactualtype-6},
		box extend=0.28}, color=red, solid,
	]
	coordinates {};
}

%

%
\end{axis}

\begin{axis}[%
boxplot/draw direction=y,
width=2.25cm,
height=2.4cm,
at={(2.484cm,0cm)},
scale only axis,
xmin=0.5,
xmax=6.5,
xlabel style={font=\color{white!15!black}, font=\footnotesize},
xlabel={Distance/m},
ymin=-8,
ymax=25,
axis x line*=bottom,
axis y line*=left,
yticklabels=\empty,
legend style={legend cell align=left, align=left, draw=white!15!black}
]

\pgfplotstableread{texfig/DistErrorData/errorBoxPlotLoc2_lin.tex}\boxdatalin
\pgfplotsinvokeforeach{0,...,5}{
	\addplot+[
	boxplot prepared from table={
		table=\boxdatalin,
		row=#1,
		lower whisker=lw,
		upper whisker=uw,
		lower quartile=lq,
		upper quartile=uq,
		median=med}, 
	boxplot prepared={
		draw position={0.8 + \plotnumofactualtype},
		box extend=0.28}, color=blue, solid,
	]
	coordinates {};
}

\pgfplotstableread{texfig/DistErrorData/errorBoxPlotLoc2_logdist.tex}\boxdatalogdist
\pgfplotsinvokeforeach{0,...,5}{
	\addplot+[
	boxplot prepared from table={
		table=\boxdatalogdist,
		row=#1,
		lower whisker=lw,
		upper whisker=uw,
		lower quartile=lq,
		upper quartile=uq,
		median=med}, 
	boxplot prepared={
		draw position={1.2 + \plotnumofactualtype-6},
		box extend=0.28}, color=red, solid,
	]
	coordinates {};
}

%

%

\end{axis}

\begin{axis}[%
boxplot/draw direction=y,
width=2.25cm,
height=2.4cm,
at={(4.968cm,0cm)},
scale only axis,
xmin=0.5,
xmax=6.5,
xlabel style={font=\color{white!15!black}, font=\footnotesize},
xlabel={Distance/m},
every outer y axis line/.append style={black},
every y tick label/.append style={font=\color{black}},
every y tick/.append style={black},
ymin=-8,
ymax=25,
axis x line*=bottom,
axis y line*=left,
yticklabels=\empty,
legend style={at={(0.45,0.55)}, anchor=south west, legend cell align=left, align=left, draw=white!15!black, font=\footnotesize, row sep=-2pt,inner sep=1pt}
]

\addlegendimage{no markers,blue}
\addlegendentry{lin}

\addlegendimage{no markers,red}
\addlegendentry{l-d}

\pgfplotstableread{texfig/DistErrorData/errorBoxPlotLoc3_lin.tex}\boxdatalin
\pgfplotsinvokeforeach{0,...,5}{
\addplot+[
boxplot prepared from table={
	table=\boxdatalin,
	row=#1,
	lower whisker=lw,
	upper whisker=uw,
	lower quartile=lq,
	upper quartile=uq,
	median=med}, 
	boxplot prepared={
	draw position={0.8 + \plotnumofactualtype},
	box extend=0.28}, color=blue, solid,
]
coordinates {};
}

\pgfplotstableread{texfig/DistErrorData/errorBoxPlotLoc3_logdist.tex}\boxdatalogdist
\pgfplotsinvokeforeach{0,...,5}{
	\addplot+[
	boxplot prepared from table={
		table=\boxdatalogdist,
		row=#1,
		lower whisker=lw,
		upper whisker=uw,
		lower quartile=lq,
		upper quartile=uq,
		median=med}, 
	boxplot prepared={
		draw position={1.2 + \plotnumofactualtype-6},
		box extend=0.28}, color=red, solid,
	]
	coordinates {};
}


%
%
%
%
%

\end{axis}
\end{tikzpicture}%

%% file: texfig/meanError_UWBdistanceLoc_1x3.tex
%
%
\definecolor{mycolor1}{rgb}{0.00000,0.44706,0.74118}%
\definecolor{mycolor2}{rgb}{0.85098,0.32549,0.09804}%
\definecolor{mycolor3}{rgb}{0.46667,0.67451,0.18824}%
\definecolor{mycolor4}{rgb}{0.92941,0.69412,0.12549}%
\begin{tikzpicture}

\node[draw=black!30, thick, rounded corners,
    minimum width=2.4cm,
    minimum height=3.0cm,anchor=south west] (pelvis) at (-0.08,-0.4) {};
\node at ($(pelvis.north) + (0,-0.25cm)$) {anechoic};

\node[draw=black!30, thick, rounded corners,
    minimum width=2.4cm,
    minimum height=3.0cm,anchor=south west] (pelvis) at (2.4,-0.4) {};
\node at ($(pelvis.north) + (0,-0.25cm)$) {corridor};

\node[draw=black!30, thick, rounded corners,
    minimum width=2.4cm,
    minimum height=3.0cm,anchor=south west] (pelvis) at (4.9,-0.4) {};
\node at ($(pelvis.north) + (0,-0.25cm)$) {corner};

\begin{axis}[%
boxplot/draw direction=y,
width=2.25cm,
height=2.4cm,
at={(0cm,0cm)},
scale only axis,
xmin=0.5,
xmax=6.5,
xlabel style={font=\color{white!15!black}, font=\footnotesize},
xlabel={Distance/m},
ymin=-8,
ymax=15,
ylabel style={font=\color{white!15!black}, font=\footnotesize},
ylabel={Distance Error / m},
axis x line*=bottom,
axis y line*=left,
legend style={legend cell align=left, align=left, draw=white!15!black}
]

\pgfplotstableread{texfig/DistErrorData/UWBerrorBoxPlotLoc1.tex}\boxdatalin
\pgfplotsinvokeforeach{0,...,5}{
	\addplot+[
	boxplot prepared from table={
		table=\boxdatalin,
		row=#1,
		lower whisker=lw,
		upper whisker=uw,
		lower quartile=lq,
		upper quartile=uq,
		median=med}, 
	boxplot prepared={
		draw position={1 + \plotnumofactualtype},
		box extend=0.35}, color=black!70, solid,
	]
	coordinates {};
}

\end{axis}

\begin{axis}[%
boxplot/draw direction=y,
width=2.25cm,
height=2.4cm,
at={(2.484cm,0cm)},
scale only axis,
xmin=0.5,
xmax=6.5,
xlabel style={font=\color{white!15!black}, font=\footnotesize},
xlabel={Distance/m},
ymin=-8,
ymax=15,
axis x line*=bottom,
axis y line*=left,
yticklabels=\empty,
legend style={legend cell align=left, align=left, draw=white!15!black}
]

\pgfplotstableread{texfig/DistErrorData/UWBerrorBoxPlotLoc2.tex}\boxdatalin
\pgfplotsinvokeforeach{0,...,5}{
	\addplot+[
	boxplot prepared from table={
		table=\boxdatalin,
		row=#1,
		lower whisker=lw,
		upper whisker=uw,
		lower quartile=lq,
		upper quartile=uq,
		median=med}, 
	boxplot prepared={
		draw position={1 + \plotnumofactualtype},
		box extend=0.35}, color=black!70, solid,
	]
	coordinates {};
}

\end{axis}

\begin{axis}[%
boxplot/draw direction=y,
width=2.25cm,
height=2.4cm,
at={(4.968cm,0cm)},
scale only axis,
xmin=0.5,
xmax=6.5,
xlabel style={font=\color{white!15!black}, font=\footnotesize},
xlabel={Distance/m},
every outer y axis line/.append style={black},
every y tick label/.append style={font=\color{black}},
every y tick/.append style={black},
ymin=-8,
ymax=15,
axis x line*=bottom,
axis y line*=left,
yticklabels=\empty,
legend style={at={(0.45,0.55)}, anchor=south west, legend cell align=left, align=left, draw=white!15!black, font=\footnotesize, row sep=-2pt,inner sep=1pt}
]

\pgfplotstableread{texfig/DistErrorData/UWBerrorBoxPlotLoc3.tex}\boxdatalin
\pgfplotsinvokeforeach{0,...,5}{
\addplot+[
boxplot prepared from table={
	table=\boxdatalin,
	row=#1,
	lower whisker=lw,
	upper whisker=uw,
	lower quartile=lq,
	upper quartile=uq,
	median=med}, 
	boxplot prepared={
	draw position={1 + \plotnumofactualtype},
	box extend=0.35}, color=black!70, solid,
]
coordinates {};
}

\end{axis}
\end{tikzpicture}%

%% file: texfig/tab_summarizedvalues.tex
\begin{tabular}{ccc|cc|cc}
Environ. & Carrying Pos. & Model & $r_p$ & $r_n$ & $F_1$ & MCC \\ 
\hline
unknown & unknown & l-d & 0.65 & 0.74 & 0.64 & 0.39\\ 
 &  & lin & 0.54 & 0.86 & 0.61 & 0.42\\ 
\hline
known & unknown & l-d & 0.68 & 0.71 & 0.64 & 0.38\\ 
 &  & lin & 0.62 & 0.79 & 0.64 & 0.42\\ 
\hline
unknown & known & l-d & 0.74 & 0.72 & 0.69 & 0.46\\ 
 &  & lin & 1.88 & 0.79 & 0.68 & 0.46\\ 
\hline
known & known & l-d & 0.75 & 0.73 & 0.70 & 0.47\\ 
 &  & lin & 0.70 & 0.77 & 0.69 & 0.47\\ 
\hline
\end{tabular}

%% file: conclusio.tex
\section{Conclusions}

To assess mobile contact tracing technology, we introduced a flexible measurement smartphone app capable to capture BLE RSSI and UWB time-of-flight distance measurements. Our experimental results reveal that BLE-based estimation of distance is sensitive to carrying positions and distance estimation. Distance estimation is not accurate but exposure detection is feasible (distances below 2.5\,m). Remarkably, multi-path propagation can reduce the effect of body shielding which may be leveraged in indoor environments where reflections from walls occur. We further showed that exposure detection can be best improved, if knowledge about the carrying position can be inferred. In future work, we collect extended measurement sets. We plan to enhance the capabilities of BLE-based distance estimation by further analysis of RSSI patterns over time to come closer to the estimation accuracy of an accurate measurement technology such as UWB.

%% file: main.bbl
\begin{thebibliography}{10}

\bibitem{nature2020}
N.~Haug, L.~Geyrhofer, A.~Londei, A.~Desvars-Larrive, V.~Loreto, B.~Pinior,
  S.~Thurnher, and P.~Klimek, ``Ranking the effectiveness of worldwide covid-19
  government interventions,'' {\em Nature Human Behaviour}, 2020.

\bibitem{martin2020demystifying}
T.~Martin, G.~Karopoulos, J.~L. Hern{\'a}ndez-Ramos, G.~Kambourakis, and I.~N.
  Fovino, ``Demystifying covid-19 digital contact tracing: A survey on
  frameworks and mobile apps,'' {\em Wireless Commun. Mobile Computing},
  pp.~1--29, Oct. 2020.

\bibitem{leith2020bleRSSI}
D.~J. Leith and S.~Farrell, ``Coronavirus contact tracing: Evaluating the
  potential of using bluetooth received signal strength for proximity
  detection,'' {\em SIGCOMM Comput. Commun. Rev.}, vol.~50, pp.~66--74, Oct.
  2020.

\bibitem{neirynck2016alternative}
D.~Neirynck, E.~Luk, and M.~McLaughlin, ``An alternative double-sided two-way
  ranging method,'' in {\em 13th Workshop Pos., Navig. Commun. (WPNC)},
  pp.~1--4, IEEE, 2016.

\bibitem{leith2020Plos}
D.~J. Leith and S.~Farrell, ``Measurement-based evaluation of google/apple
  exposure notification api for proximity detection in a light-rail tram,''
  {\em PLoS ONE}, no.~9, 2020.

\bibitem{benkic2008using}
K.~Benkic, M.~Malajner, P.~Planinsic, and Z.~Cucej, ``Using rssi value for
  distance estimation in wireless sensor networks based on zigbee,'' in {\em
  2008 15th International Conference on Systems, Signals and Image Processing},
  pp.~303--306, IEEE, 2008.

\bibitem{zhao2014does}
X.~Zhao, Z.~Xiao, A.~Markham, N.~Trigoni, and Y.~Ren, ``Does {BTLE} measure up
  against {WiFi}? a comparison of indoor location performance,'' in {\em
  European Wireless 2014; 20th European Wireless Conference}, pp.~1--6, VDE,
  2014.

\bibitem{touvat2014indoor}
F.~Touvat, J.~Poujaud, and N.~Noury, ``Indoor localization with wearable rf
  devices in 868mhz and 2.4 ghz bands,'' in {\em 2014 IEEE 16th International
  Conference on e-Health Networking, Applications and Services (Healthcom)},
  pp.~136--139, IEEE, 2014.

\bibitem{sadowski2018rssi}
S.~Sadowski and P.~Spachos, ``{RSSI}-based indoor localization with the
  internet of things,'' {\em IEEE Access}, vol.~6, pp.~30149--30161, 2018.

\bibitem{MITPACTdata2020}
C.~Corey, ``{PACT} datasets and evaluation website.''
  \url{https://github.com/mitll/PACT}, 2020.

\bibitem{DP3Tdata2020}
DP-3T, ``Bluetooth measurements.''
  \url{https://github.com/DP-3T/bt-measurements}, 2020.

\bibitem{etzlinger2020UWB}
B.~Etzlinger, A.~Ganh\"r, J.~Karoliny, R.~H\"uttner, and A.~Springer, ``{WSN}
  implementation of cooperative localization,'' in {\em Proc. IEEE MTT-S Int.
  Conf. Microwaves and Intelligent Mobility}, pp.~1--4, IEEE, 2020.
\newblock to appear.

\bibitem{etzlinger2018synchronization}
B.~Etzlinger, H.~Wymeersch, {\em et~al.}, ``Synchronization and localization in
  wireless networks,'' {\em Foundations and Trends{\textregistered} in Signal
  Processing}, vol.~12, no.~1, pp.~1--106, 2018.

\bibitem{gezici2005localization}
S.~Gezici, Z.~Tian, G.~B. Giannakis, H.~Kobayashi, A.~F. Molisch, H.~V. Poor,
  and Z.~Sahinoglu, ``Localization via ultra-wideband radios: a look at
  positioning aspects for future sensor networks,'' {\em IEEE signal processing
  magazine}, vol.~22, no.~4, pp.~70--84, 2005.

\bibitem{gustafson2015statistical}
C.~Gustafson, T.~Abbas, D.~Bolin, and F.~Tufvesson, ``Statistical modeling and
  estimation of censored pathloss data,'' {\em IEEE Wireless Communications
  Letters}, vol.~4, no.~5, pp.~569--572, 2015.

\end{thebibliography}
